\documentclass{appolb}
\usepackage{graphicx,color}
\usepackage{mathrsfs,slashed}

\def\bea{\begin{eqnarray}}
\def\eea{\end{eqnarray}}
\def\be{\begin{equation}}
\def\ee{\end{equation}}

\begin{document}
\title{Generalized Beth-Uhlenbeck approach to the \\ equation of state for quark-hadron matter%
\thanks{Talk at "Critical Point and Onset of Deconfinement", Wroc{\l}aw, 30.05.-04.06.2016}%
}
\author{D. Blaschke$^{a,b,c}$, A. Dubinin$^a$, L. Turko$^{a}$
\address{$^{a}$Institute of Theoretical Physics, University of Wroclaw, 50-204 Wroclaw, Poland
\\$^{b}$Bogoliubov Laboratory of Theoretical Physics, JINR, 141980 Dubna, Russia
\\$^{c}$National Research Nuclear University (MEPhI), 115409 Moscow, Russia
}
}
\maketitle

\begin{abstract}
A unified equation of state for quark-hadron matter is presented in the generalized Beth-Uhlenbeck form. 
It follows from a $\Phi-$derivable approach to the thermodynamic potential where the ansatz for the 
$\Phi$ functional contains all  2PI diagrams at two-loop order formed with quark cluster Green's functions for quark, diquark, meson and baryon propagators.  
We present numerical results using an effective model for the generic behaviour of hadron masses and phase shifts at finite temperature which shares basic features with recent developments within the PNJL model for correlations in quark matter.
We obtain the transition between a hadron resonance gas phase and the quark gluon plasma where the Mott dissociation of hadrons is encoded in the hadronic phase shifts.
The resulting thermodynamics is in very good agreement with recent lattice QCD simulations.
\end{abstract}


\section {Introduction} \label{Intrd}

The aim of this contribution is to provide a unified approach to the hot QCD equation of state in agreement with recent independent lattice QCD (LQCD) simulations \cite{Borsanyi:2013bia,Bazavov:2014pvz,Borsanyi:2010bp}, well reproducing both limits, the hadron resonance gas at low temperatures and the quark-gluon plasma (QGP) with perturbative QCD corrections at high temperatures and describing  the crossover between both by the Mott mechanism.

The quark and gluon degrees of freedom are described within an effective
meanfield theory, the Polyakov-loop improved Nambu--Jona-Lasinio model.
The in-medium effect responsible for the hadron-to-quark matter
phase transition is the lowering of the quark masses in the chiral restoration
transition which itself is a result of the behaviour of the chiral condensate.
We take the solution for the temperature dependent chiral condensate as an input
from full LQCD simulations and focus on a description of the hadron sector embodying
the Mott effect of hadron dissociation within a generalized Beth-Uhlenbeck approach
based on in-medium hadron phase shifts in accordance with the Levinson theorem.

We postulate a generic behaviour of the scattering
phase shifts in the hadronic channels which are temperature dependent and embody the
main consequence of chiral symmetry restoration in the quark sector: the
lowering of the thresholds for the two- and three-quark scattering state
continuous spectrum which triggers the transformation of hadronic bound states
to resonances in the scattering continuum.
The phase shift model is in accordance with the Levinson theorem which results
in the vanishing of hadronic contributions to the thermodynamics at high
temperatures.

\section{$\Phi-$derivable and generalized Beth-Uhlenbeck approach}

\begin{figure}[!htb]
\parbox{0.2\textwidth} {
$\Phi\left[\left\{G_i \right\} \right]=$
}
\parbox{0.02\textwidth}{$\frac{1}{2}$}
\parbox{0.12\textwidth}{
\includegraphics[width=0.12\textwidth]{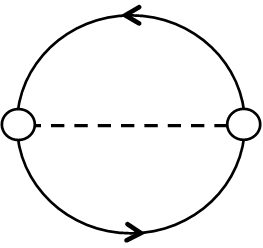}
}
\parbox{0.04\textwidth}{$+\frac{1}{2}$}
\parbox{0.12\textwidth}{
\includegraphics[width=0.12\textwidth]{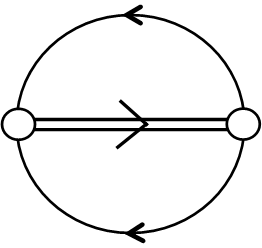}
}
\parbox{0.02\textwidth}{$+$
}
\parbox{0.12\textwidth}{
\includegraphics[width=0.12\textwidth]{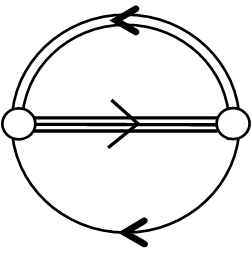}
}
\\
\parbox{0.15\textwidth} {
\bea
\Sigma_q=\frac{\delta \Phi}{\delta G_q}=\nonumber
\eea
}\hspace{2mm}
\parbox{0.12\textwidth}{
\includegraphics[width=0.12\textwidth]{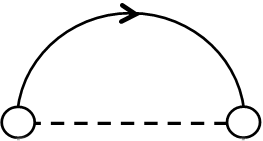}
}
\parbox{0.01\textwidth}{$+$
}
\parbox{0.12\textwidth}{
\includegraphics[width=0.12\textwidth]{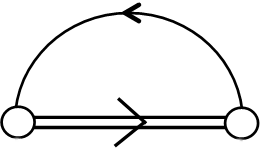}
}
\parbox{0.01\textwidth}{$+$
}
\parbox{0.12\textwidth}{
\includegraphics[width=0.12\textwidth]{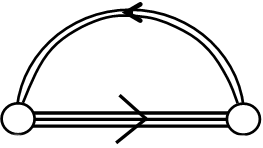}
}
\caption{
Upper line: Diagram choice for the $\Phi$ functional in the non-perturbative sector of low-energy QCD where strong correlations in the mesonic (dashed line), diquark (double line) and baryonic (triple line) channels are present.
Lower line: Quark selfenergies corresponding to the $\Phi$ functional of the upper line.
\label{fig:Sigma}}
\end{figure}

For the thermodynamic potential of quark-hadron matter we suggest to employ an ansatz in the spirit of 
the $\Phi-$derivable approach \cite{Baym:1961zz,Baym:1962sx}, 
where besides the full propagator for quarks (q) also those for the diquarks (d), mesons (M) and baryons (B) as quark composites appear
\bea
\label{Omega}
\Omega = \sum_{i=q,d,M,B} \frac{{c_i}}{2}\left[{\rm Tr} \ln(G_i^{-1}) - {\rm Tr} (\Sigma_i G_i)  \right]
+ \Phi\left[\{G_i\} \right] 
+ \mathscr{U}[\phi;T]
+ \Omega_{\rm pert}.
\eea
The approximation for the 2-particle irreducible $\Phi$ functional contains all two-loop diagrams of the "sunset" type, and it generates the corresponding selfenergies. 
In Fig.~\ref{fig:Sigma} we show this diagram choice and the resulting quark selfenergies.
This is a cluster virial expansion for quark matter \cite{Blaschke:2015bxa} analogous to the one known for 
nuclear matter  \cite{Ropke:2012qv,Typel:2009sy} leading to a generalized Beth-Uhlenbeck equation of state \cite{Schmidt:1990rs}. For details, see \cite{Blaschke:2016hzu}.

\noindent
For the total pressure of the model, we have
\begin{eqnarray}
P_{\rm total}(T) = \sum_{i = M,B}{P_i(T)}+P_{\rm PNJL}^*(T)+P_{\rm pert}(T)~,
\label{eq:15}
\end{eqnarray}
where the first term describes a Mott hadron resonance gas (MHRG) as  ideal mixture of hadronic bound and scattering states in the channels $i$ that follow ideal gas-like behavior  
\bea
\label{P-i}
P_i(T) &=& T\ n_i(T)~,~~ i=\{M\}, \{B\},
\eea
with the generalized  scalar density in the hadronic channel $i$, 
\bea
n_i(T)&=&	\int_0^\infty \frac{d\hat{M}}{\pi}n_{s,i}(\hat{M})
	 [\delta_i(\hat{M}^2)- \frac{1}{2}\sin 2\delta_i(\hat{M}^2)]. 
\label{P_HRG}
\eea
This separability of the hadronic contributions holds at two-loop order for the $\Phi$ functional
diagram choice \cite{Blaizot:2000fc,Vanderheyden:1998ph}. 
The Mott dissociation effect is encoded  
in the temperature-dependent hadron phase shifts $\delta_i(\hat{M}^2)$ in (\ref{P_HRG}).
The sinus term in (\ref{P_HRG}) marks a difference with the traditional Beth-Uhlenbeck approach 
\cite{Beth:1936zz,Hufner:1994ma,Wergieluk:2012gd,Blaschke:2013zaa}
that does not have this term which leads to a "squared Breit-Wigner" spectral shape instead of a Breit-Wigner one
\cite{Vanderheyden:1998ph,Morozov:2009}. 
The scalar density of a hadronic degree of freedom with mass $M$ is
\bea
n_{s,i}(\hat{M})&=& d_iT^3 \int_0^\infty\frac{d\hat{p}~\hat{p}^2}{2\pi^2}
	\frac{\hat{M}}{\sqrt{\hat{p}^2+\hat{M}^2}}f_i(\sqrt{\hat{p}^2+\hat{M}^2})~,
	\nonumber
\eea
with $d_i$ being the degeneracy of the state $i$ and 
$f_i(x)=[{\rm e}^{x}- {c_i}]^{-1}$
the corresponding distribution function; $\hat{M}=M/T$, $\hat{p}=p/T$.

The underlying quark and gluon thermodynamics is divided into a perturbative contribution 
$P_{\rm pert}(T)$ which is 
treated as virial correction in two-loop order following Ref.~\cite{Turko:2011gw} and a nonperturbative   
part described within a PNJL model in the form
\bea
\label{P_PNJL}
P_{\rm PNJL}^*(T) = P_{\rm FG}^*(T) +  \mathscr{U}[\phi;T]~,
\eea
where  the Polyakov-loop potential $\mathscr{U}[\phi;T]$ takes into account the nonperturbative gluon background in a meanfield approximation using the polynomial fit of Ref.~\cite{Ratti:2005jh}.
The asterix denotes that we go beyond the standard meanfield level and introduce a quasiparticle
picture
\begin{eqnarray}
P_{\rm FG}^*(T) = 4N_c \sum_{q=u,d,s} \int\frac{dp~p^2}{2 \pi^2}\int\frac{d\omega}{\pi}
f_{\phi}(\omega) 
\left\{\delta_q(\omega) -\frac{1}{2}\sin[2\delta_q(\omega)] \right\},
\label{eq:P-FG*}
\end{eqnarray}
where the generalized Fermi distribution function of the PNJL model for the case of vanishing
quark chemical potential considered here is defined as
$f_{\phi}(\omega)=[\phi(1 + 2 Y) Y + Y^3]/[1+3\phi(1+ Y)Y + Y^3]$~,
with $Y=\exp(-\omega/T)$.
The quark phase shift due to the scattering off hadrons is taken as
\bea
\delta_q(\omega) = \pi H(\omega,E_p,\gamma)~,~~~
H(x,y,z)= {1}/{2} + ({1}/{\pi})\arctan \left[ ({x-y})/{z}\right],
\eea
where $E_p=\sqrt{p^2+m_q^2}$ is the quark dispersion relation ($q=u,d,s$) and the parameter $\gamma$
stands for the collisional broadening \cite{Friesen:2013bta}
\begin{equation}
\gamma(T) = \sum_{i = M,B} \sigma ~n_{i}(T)~,
\end{equation}
where for the cross section we adopt a universal value of $\sigma=35$ mb that is guided by the asymptotic
nucleon-nucleon cross section. 
For the case of vanishing width parameter the usual Fermi gas expression for the quark pressure of the PNJL model is reproduced.

\section{Temperature dependent quark and hadron spectrum}

We are solving the standard gap equation for the traced Polyakov loop $\phi(T)$  with
the quark masses $m_l(T)$ and $m_s(T)$ as an input, where $l=u,d$ denotes the degenerate light quark
flavors. 
The temperature dependence of the quark masses is obtained using LQCD data for the behaviour
of the continuum extrapolated chiral condensate $\Delta_{l,s}(T)$ \cite{Borsanyi:2010bp}. 
In such a way it is possible to go beyond the meanfield and include effects of hadronic resonances consistently.
For the temperature dependence of light quark mass $m(T)$ we assume
\begin{equation}
\label{m_light}
m(T)=[m(0)-m_0]\Delta_{l,s}(T)+m_0~,
\end{equation}
with $m_0=5.5$ MeV
and for the strange quark mass we adopt
$m_s(T) = m(T) + m_s - m_0~,$
with $m_s = 100$ MeV.

For the hadron masses $M_i(T)$ and widths $\Gamma_i(T)$ we make the ansatz
\bea
\label{spectrum}
M_i(T) &=& M_i(0) + \Gamma_i(T)~~,\\
\Gamma_i(T) &=& \sqrt{a ~(T-T_{\rm Mott, i})+b~(T-T_{\rm Mott, i})^2} ~ \Theta(T-T_{\rm Mott, i})~,
\eea
where $M_i(0)$ are the hadron masses according to the particle data group, for the 
parameters we choose $a=2.5$ GeV and $b=8.0$~\cite{Hufner:1996pq}. 
For hadrons that are unstable at $T=0$ already we use the linear fit ($b=0$), with negative Mott temperature.

\begin{figure}
\includegraphics[width=0.52\textwidth]{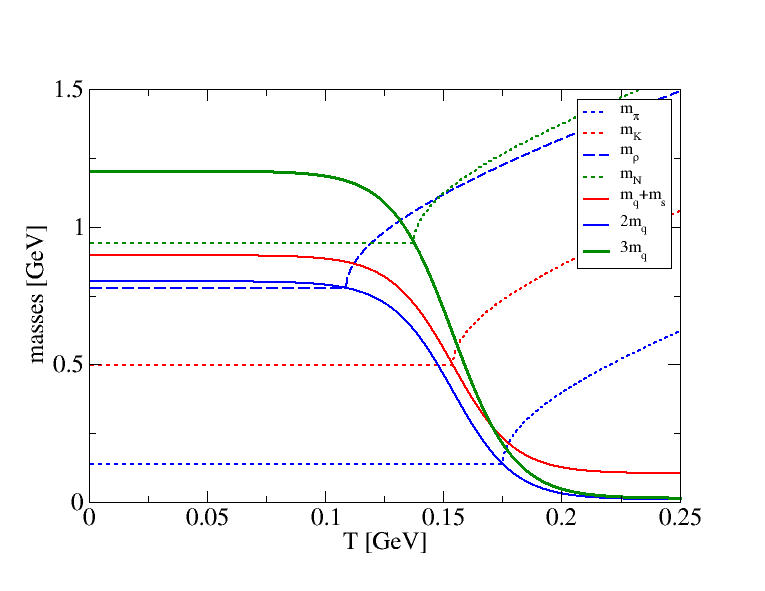}
\includegraphics[width=0.48\textwidth,angle=0]{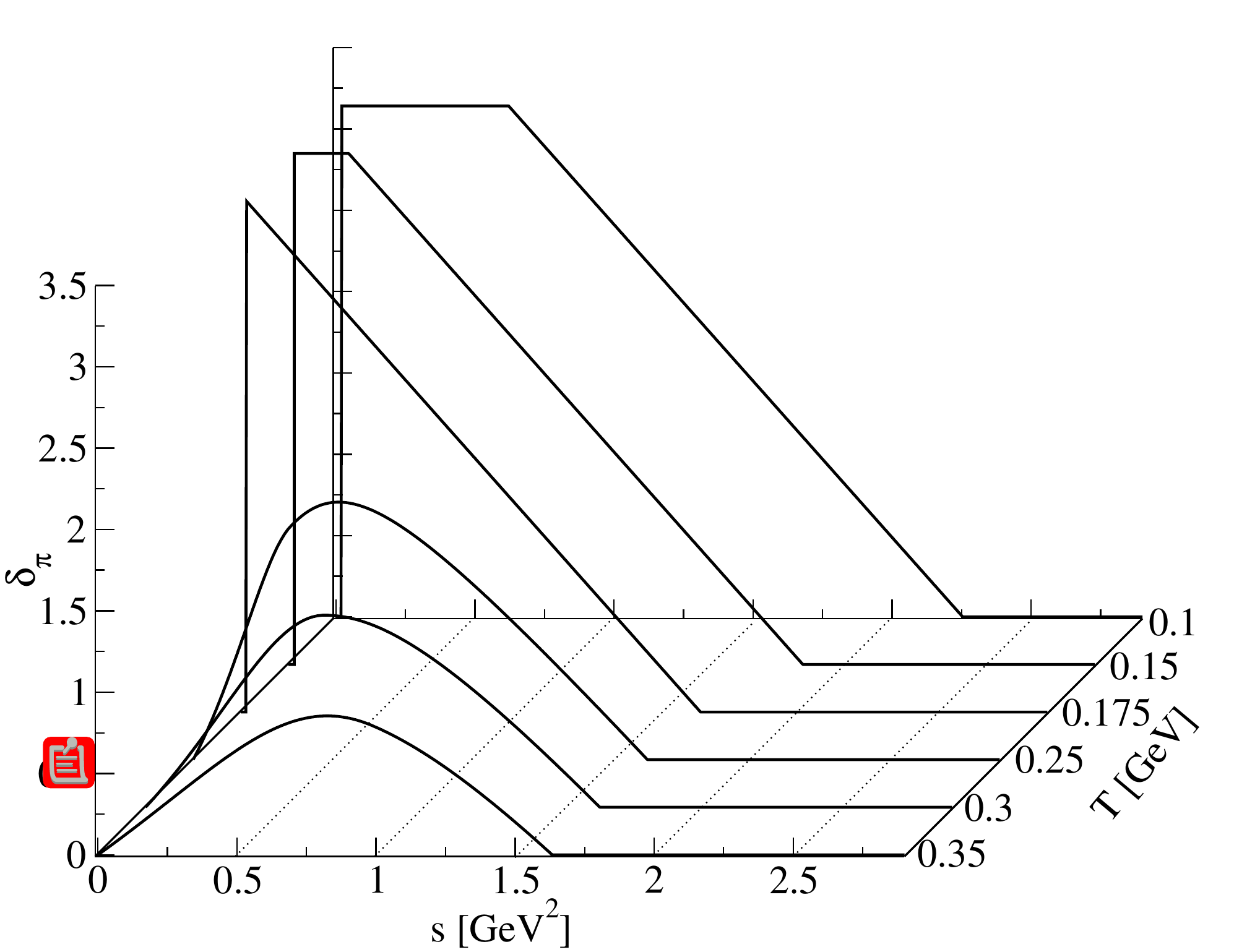}
\caption{Left panel: 
Temperature dependence of the light hadron masses and the corresponding 2-quark and 
3-quark continuum thresholds (solid lines).
Right panel: 
Temperature and s-dependence of the pion phase shift according to the generic hadron phase shift model of Eq.~(\ref{phaseshift}).
\label{fig:spectrum}}
\end{figure}

The Mott temperatures $T_{\rm Mott,i}$ can be determined from the condition
\bea
\label{Mott}
M_i(T_{\rm Mott,i})&=& m_{\rm thr, i}(T_{\rm Mott,i})~,
\eea
where the temperature dependent continuum threshold for 
a hadron species $i$ containing $N_i$ valence quarks is 
determined by the temperature dependent quark masses 
via 
\bea
\label{Mott}
m_{\rm thr, i}(T)&=&(N_i-N_s)m(T)+N_s m_s(T)~~,
\eea
where $N_s=0,1,\dots , N_i$ is the number of strange quarks in hadron $i$ with $N_i=2$ for mesons 
($i=M$) and $N_i=3$  for baryons ($i=B$).
The resulting mass spectrum is shown in the left panel of Fig.~\ref{fig:spectrum}.

For the hadronic phase shifts we adopt the generic function
\bea
\label{phaseshift}
\delta_i(\hat{s})&=&\pi\ F\left({\hat{s}}/{\hat{\Gamma}_i^2}\right)
H(\hat{s}, \hat{M}_i^2,\hat{M}_i\hat{\Gamma}_i) \left\{
\Theta(\hat{s}_{\rm thr,i} - \hat{s}) 
\right. \nonumber\\
&&\left. +\left[\frac{\hat{s}_{\rm max,i}-\hat{s}}{\hat{s}_{\rm max,i}-\hat{s}_{\rm thr,i}} \right] 
 \Theta(\hat{s}-\hat{s}_{\rm thr,i}) \Theta(\hat{s}_{\rm max,i} - \hat{s}) 
\right\},
\eea
where 
$\hat{s}=s/T^2$, $\hat{M}_i=M_i/T$, $\hat{\Gamma}_i=\Gamma_i/T$.
The auxiliary function
$F(x)= [\sin(x)\Theta(\pi/2-x) + \Theta(x - \pi/2)]$
has been introduced to ensure that the phase shift at $s=0$ is always zero, even at higher
temperatures, where large values of the width parameter in the Breit-Wigner like ansatz could 
otherwise spoil this constraint. 
Above the continuum threshold $s_{\rm thr,i}=m_{\rm thr,i}^2$ the phase shift drops towards zero which is reached at $s_{\rm max,i}=s_{\rm thr,i}+N_i^2\Lambda^2$, where $\Lambda$ is the range of the nonpertubative interaction in momentum space.
In the right panel of Fig.~\ref{fig:spectrum} we illustrate the temperature and energy dependence encoded in this formula for the case of the pion.  
Note the similarity to the prototype phase shift in \cite{Dashen:1974yy} that fulfils the basic requirement of the Levinson theorem while encoding the behaviour of a resonance.

\section{Results and conclusions}

\begin{figure}[!htb]
\includegraphics[width=0.52\textwidth,angle=0]{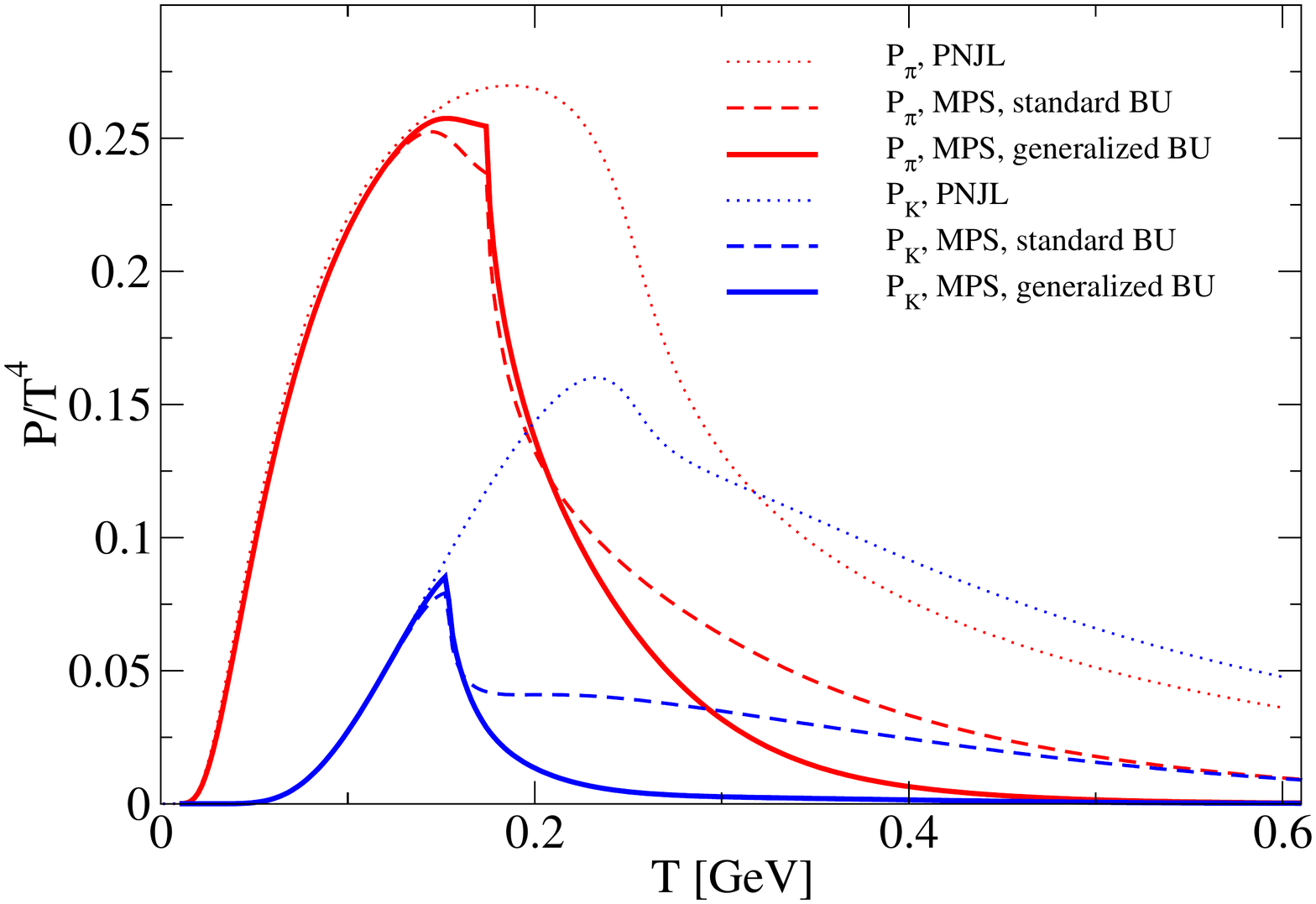}
\includegraphics[width=0.52\textwidth,angle=0]{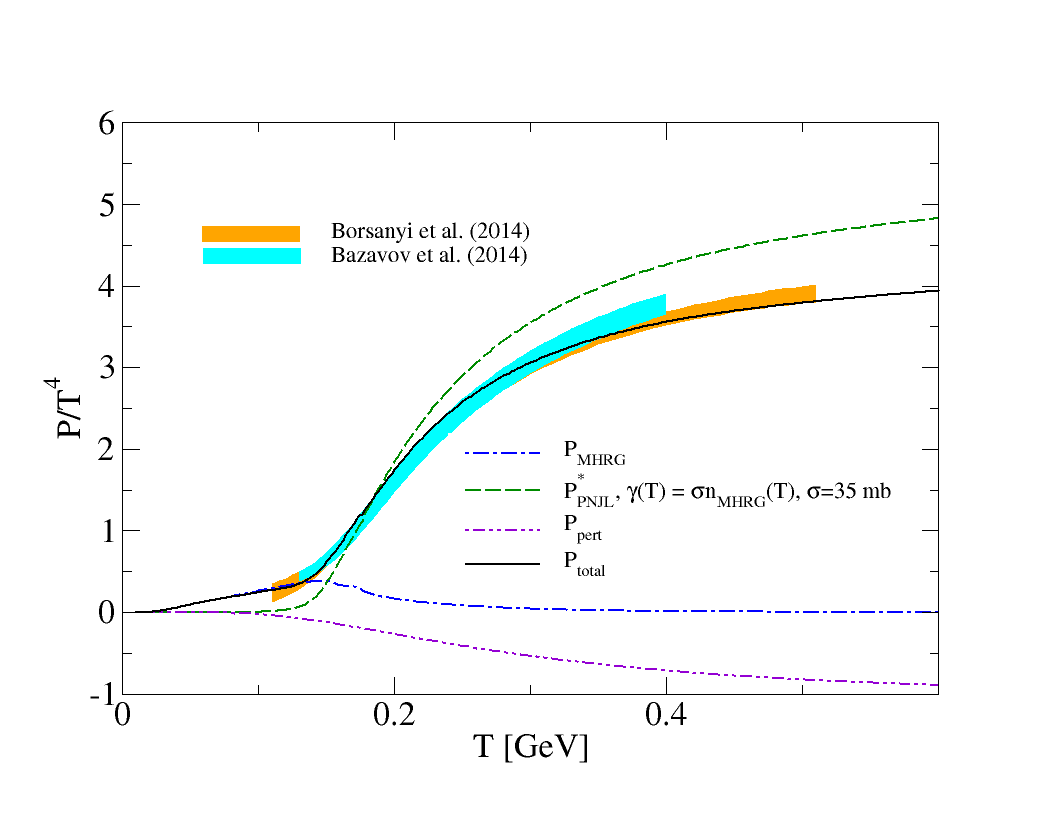}
\caption{
Left panel: Temperature dependence of the pressure for the pion and kaon component of the MHRG.
We compare the results for the standard PNJL model (dotted lines) where the Mott temperature is 
around $250$ MeV for both species with the model phase shift results in the standard Beth-Uhlenbeck form (dashed lines) and the generalized Beth-Uhlenbeck form (solid lines).
Right panel:
Temperature dependence of the total pressure of the present model (black solid line) compared to the LQCD results of Ref.~\cite{Borsanyi:2013bia,Bazavov:2014pvz}.
For comparison, the hadron (blue dash-dotted line), quark-gluon (green dashed line) and perturbative QCD contributions (violet dash-double-dotted line) are shown.
 \label{pressure-PiK}
}
\end{figure}

We are now in the situation to discuss the results for the QCD thermodynamics as captured by our model
in Eq.~(\ref{eq:15}), where the different components are defined above.
In the left panel of Fig.~\ref{pressure-PiK} we show the result for the temperature dependence of the partial pressure of pions and kaons obtained in the above MHRG model, within different approximations. 

In the right panel of Fig.~\ref{pressure-PiK} we compare the total pressure of the present model (P$_{\rm total}$) with LQCD data from Ref.~\cite{Borsanyi:2013bia,Bazavov:2014pvz}. We show also the partial pressure contributions from the hadronic correlations (P$_{\rm MHRG}$), the nonperturbative (P$_{\rm PNJL}^*$) and the perturbative (P$_{\rm pert}$) quark-gluon models. 

With the present schematic model, we can quantify the transition temperature from hadron dominance to quark-gluon dominance at $T_{\rm trans}=156$ MeV, where both partial pressures are equal.
This falls in the range of the pseudocritical transition temperature found in LQCD simulations, which
is also reflected in the temperature dependence of the chiral condensate and our fit thereof.
At the same time the scaled pion and kaon pressure components both peak at the temperature of
$T=153$ MeV, related to the Mott dissociation of these states.
Furthermore one can identify the asymptotic limits, the pion gas limit at low temperatures $T < 100$ MeV and the quark-gluon plasma limit at high temperatures above $T\sim 250$ MeV.

Note that we have found that the account of both effects, the collisional width due to quark-hadron scattering in $P_{\rm FG}^*(T)$ and the virial corrections by parton rescattering to order $\alpha_s$ in $P_{\rm pert}(T)$ together are important for obtaining an excellent approximation to QCD thermodynamics in the whole domain of temperatures sampled by the recent LQCD results. 
Without these contributions there remain strong discrepancies, as demonstrated recently in Ref.~\cite{Torres-Rincon:2016ahl}.
			
In this work an effective model is constructed which is capable of reproducing basic physical characteristics of the hadron resonance gas at low temperatures and embody the crucial effect of hadron dissociation by the Mott effect.
The generalized Beth-Uhlenbeck form of the partial pressures is constructed for each hadronic channel. Numerical results show that the simplifying ansatz for the temperature dependence of both, the mass spectrum and the phase shifts of hadronic channels give results in quantitative agreement with recent ones from LQCD \cite{Borsanyi:2010cj}.
To achieve this it was essential to realize a calculational scheme that is inspired by the $\Phi-$derivable approach of Baym and Kadanoff.
Due to the confining property of QCD it is of crucial importance to take into account the strong contributions of the hadron resonance gas components to the quark degrees of freedom which constitute them.
The generalized Beth-Uhlenbeck scheme proves essential to overcome discrepancies in the quantitative
modeling of LQCD thermodynamics that existed in a previous version of this model
\cite{Blaschke:2015nma}. 
The hadronic Mott effect provides a proper understanding of hadronic dissociation phenomena and can be formulated within an extended PNJL model augmented by virial corrections from partonic scattering at two-loop order. 

The present model describes the QCD thermodynamics in accordance with state-of-the-art 
LQCD simulations and thus provides an interpretation of the latter as well as a basis for modeling the full QCD phase diagram when extending it to finite chemical potentials currently inaccessible to LQCD due to the sign problem. 

\subsection*{Acknowledgement}
Support by NCN under contract No. UMO-2011/02/A/ST2/00306 (D.B. and A.S.) and contract No. UMO-2014/15/B/ST2/03752 (L.T.) is gratefully acknowledged.
D.B. was supported in part by the MEPhI Academic Excellence Project under contract No. 02.a03.21.0005.

{}

\end{document}